\documentclass{amsart}
\usepackage{graphicx} 

\usepackage{
  amscd,
  amsfonts,
  amsmath,
  amssymb,
  amsbsy,
  amsthm,
  bm, 
  dsfont,
  cancel, 
  latexsym,
  cases,
  setspace,
  mathrsfs,
  mathtools,
  graphicx,
  placeins,
  xcolor,
  xargs,
  lineno
}

\usepackage[foot]{amsaddr}
\usepackage[utf8]{inputenc}
\usepackage[shortlabels]{enumitem}

\usepackage{
  mathtools,
  booktabs,
  caption,
  etoolbox,
  enumitem,
  float,
  latexsym,
  lipsum,
  placeins,
  mathrsfs,
  setspace,
  subcaption,
  url,
  graphicx,
  xcolor
}

\usepackage{caption, subcaption}

\makeatletter
\def\@setthanks{\vspace{-\baselineskip}\def\thanks##1{\@par##1\@addpunct.}\thankses}
\makeatother

\usepackage{rotating}
\usepackage{lscape}
\newcommand{\bland}{\begin{landscape}}
\newcommand{\eland}{\end{landscape}}

\newcommand{\bburl}[1]{\textcolor{blue}{\url{#1}}}

\usepackage[
  hyperref=true,
  url=false,
  isbn=false,
  backref=true,
  maxcitenames=3,
  backend=biber,
  style=authoryear,
  citestyle=authoryear,
  natbib=true]{biblatex}

\usepackage[colorlinks=true]{hyperref}

\newtheorem{remark}{Remark}

\definecolor{maroon}{rgb}{0.5, 0.0, 0.0}
\hypersetup{breaklinks=true,
            bookmarks=true,
            pdfauthor={Apoorva Lal},
             pdfkeywords = {},
            colorlinks=true,
            citecolor=maroon,
            urlcolor=blue,
            linkcolor=blue,
            pdfborder={0 0 0}}
\makeatletter
\def\section{\@startsection{section}{1}%
      \z@{.7\linespacing\@plus\linespacing}{.5\linespacing}%
      {\normalfont\Large\bfseries\centering }}

\def\sectionL{\@startsection{section}{1}%
      \z@{.7\linespacing\@plus\linespacing}{.5\linespacing}%
      {\normalfont\Large\bfseries}}
\makeatother

\patchcmd{\subsection}{\bfseries}{\bfseries\large}{}{}
\patchcmd{\subsubsection}{\itshape}{\bfseries}{}{}

\makeatletter
\def\paragraph{\@startsection{paragraph}{4}%
  \z@\z@{-\fontdimen2\font}%
  {\sffamily \bfseries }}
\makeatother

\usepackage{tikz}

\usetikzlibrary{graphs, quotes}

\newcommand{\tauhat}{\hat{\tau}}

\newcommand{\E}{\mathbb{E}}
\newcommand{\R}{\mathbb{R}}
\newcommand{\Var}{\text{Var}}
\newcommand{\Cov}{\text{Cov}}
\newcommand{\plim}{\text{plim}}
\newcommand{\tra}{\prime}

\newcommand{\AMSE}{\mathrm{AMSE}}

\newcommand{\matr}[1]{#1}
\newcommand{\vect}[1]{#1}

\usetikzlibrary{arrows.meta, positioning, shapes, fit}

\usepackage{XCharter}

\usepackage[margin=1in]{geometry}
\usepackage[
    disable 
]{todonotes}

\usepackage{comment}

\makeatletter
\providecommand\@dotsep{5}
\def\listtodoname{}
\def\listoftodos{\@starttoc{tdo}\listtodoname}
\makeatother

\newcommand{\hl}{\noindent\rule{\textwidth}{1pt}}

\title[]{
    Long-term Causal Inference with Many Noisy Proxies
}

\author{Apoorva Lal}
\address{Lal: Independent Researcher, work done while at amazon}
\email{lal.apoorva@gmail.com}
\author{Guido Imbens}
\address{Imbens: Stanford University, Amazon}
\email{imbens@stanford.edu}
\author{Peter Hull}
\address{Hull: Brown University, Amazon}
\email{peter\_hull@Brown.edu}
\date{\today}

\begin{document}


\maketitle

\begin{abstract}
    We propose a method for estimating long-term treatment effects with many short-term proxy outcomes: a central challenge when experimenting on digital platforms. We formalize this challenge as a latent variable problem where observed proxies are noisy measures of a low-dimensional set of unobserved surrogates that mediate treatment effects. Through theoretical analysis and simulations, we demonstrate that regularized regression methods substantially outperform naive proxy selection. We show in particular that the bias of Ridge regression decreases as more proxies are added, with closed-form expressions for the bias-variance tradeoff. We illustrate our method with an empirical application to the California GAIN experiment.
\end{abstract}

\hl

\section{Introduction}

Estimating the long-term effects of interventions is a central
challenge for firms operating in the digital economy \citep{gupta2019top}. While randomized controlled trials (RCTs) can cleanly identify the short-term impacts of a new product feature, marketing campaign, or pricing strategy, their timeframe is often too short to capture long-term outcomes of ultimate interest---such as customer retention or lifetime value. Consequently, firms usually rely on short-term proxies which are measured within the experimental window and are believed to be at least somewhat predictive of long-term outcomes. As these short-term measures are often easy to measure in modern online platforms, the number of possible short-run proxies can be quite large. 

This paper proposes methods for using many short-term proxy outcomes to estimate long-term treatment effects, building on a long literature on surrogate analysis. Classic surrogate methods, arising from surrogate-endpoint analysis in medical trials, assume the existence of a single or small handful of proxies which fully mediate long-run effects \parencite[e.g.,][]{prentice1989surrogate}. More modern analyses allow the number of surrogates to grow large and complex while still maintaining the ``surrogacy criterion'' of complete mediation \parencite[e.g.,][]{athey2019surrogate,
chen2023semiparametric,Kallus2020-kh}. In many real-world settings, however, complete surrogacy may be implausible. A more realistic assumption may be that the many available short-run outcomes---such as on clicks, add-to-cards, session duration, and other forms of online activity at an e-commerce platform experimenting with a new recommendation algorithm---are noisy measures of some underlying set of latent surrogates which are never observed. We study this setting with many noisy proxies. 


We develop our method with a simple economic model. A binary
treatment $W$ (e.g., exposure to a new recommendation algorithm or marketing
intervention) is randomized to customers by a firm and potentially affects each customer's unobserved
state $S$ (e.g., their engagement or satisfaction). This latent vector fully mediates the effects of $W$ on a long-term outcome of interest (e.g., future engagement or spending). Rather than observing $S$, however, the econometrician sees many short-term proxies $P$ which are noisy measures of $S$. Crucially, the number of noisy proxies in $P$ is allowed to be much larger than the number of  surrogates. 



In this setting, we propose using regularized regression methods to construct a surrogate index from the noisy proxies. By systematically shrinking the weight put on different proxies, methods like Ridge, Lasso, or partial least squares (PLS) can exploit the proxy structure---effectively filtering out noise to recover a reliable measure of the underlying latent surrogate. We show both theoretically and empirically how such methods dominate other common approaches in this setting, such as the naive approach of selecting and using the proxies that are most correlated with the long-term outcome in a conventional surrogate analysis. 


Specifically, our contributions here are fourfold. First, we formalize the problem of surrogate selection with many noisy proxies as one of latent variable regression with measurement error. Second, we provide a theoretical analysis of Ridge regression in this setting, deriving a simple, interpretable form for the asymptotic bias that clarifies the role of regularization and the
number of available proxies. Third, through extensive numerical
simulations, we demonstrate that regularized estimators (Ridge, Lasso, and PLS) consistently and substantially outperform naive screening methods, particularly in the challenging high-dimensional settings
that motivate our work. Finally, we find that the approach has favourable properties in estimating long-term effects on the California GAIN experiments \parencite{hotz2006evaluating}, and to  demonstrate the use of the method with varying numbers of surrogates,  we provide an interactive application that can be accessed at \href{https://linearsurrogateindex.streamlit.app/}{https://linearsurrogateindex.streamlit.app/}.



\section{Setup}

We consider a setting where an intervention's effect on a long-term outcome is mediated
by a low-dimensional unobserved vector, which we call the state or
surrogate. We have access to two distinct data sources: an
experimental dataset for estimating short-term effects on many noisy proxies of the surrogate and an
observational dataset containing long-term outcomes and these proxies.

\subsection{The Causal Model}

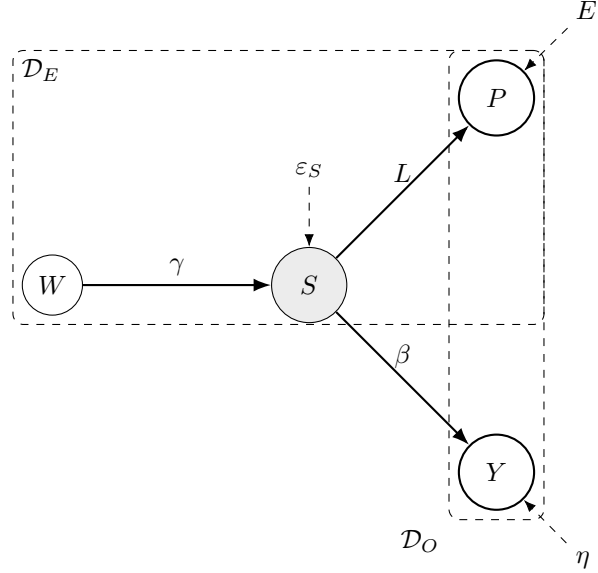
\begin{figure}
\centering
\begin{tikzpicture}[
    node distance=2.5cm,
    latent/.style={circle, draw, minimum size=1cm, fill=gray!15},
    observed/.style={circle, draw, thick, minimum size=1cm},
    treatment/.style={circle, draw, minimum width=0.8cm, minimum height=0.8cm}
]

\node[treatment] (W) {$W$};
\node[latent, right=of W] (S) {$S$};
\node[observed, above right=of S] (P) {$P$};
\node[observed, below right=of S] (Y) {$Y$};

\draw[-{Latex}, thick] (W) -- node[midway, above] {$\gamma$} 
                            node[midway, below, font=\footnotesize\itshape] { } (S);
                            
\draw[-{Latex}, thick] (S) -- node[midway, above] {$L$} 
                            node[midway, below, font=\footnotesize\itshape] { } (P);
                            
\draw[-{Latex}, thick] (S) -- node[midway, above] {$\beta$} 
                            node[midway, below, font=\footnotesize\itshape] { } (Y);

\node[above=0.8cm of S] (eS) {$\varepsilon_S$};
\node[above right=0.8cm of P] (eP) {$E$};
\node[below right=0.8cm of Y] (eY) {$\eta$};

\draw[-{Latex}, dashed] (eS) -- (S);
\draw[-{Latex}, dashed] (eP) -- (P);
\draw[-{Latex}, dashed] (eY) -- (Y);

\node[draw, dashed, rounded corners, fit=(W) (P), label={[anchor=north
west]north west:$\mathcal{D}_E$}] {};
\node[draw, dashed, rounded corners, fit=(P) (Y), label={[anchor=north
east]south west:$\mathcal{D}_O$}] {};


\end{tikzpicture}
\caption{Graphical representation of the causal model with unobserved
surrogate, and two partially overlapping data sources. The gray node
denotes that $S$ is unobserved.}
\label{fig:dag_prox}
\end{figure}

Our data generating process is based on the causal graph in
figure \ref{fig:dag_prox}. Let $W$ be a binary treatment, $S \in \R^p$
be the unobserved p-dimensional state  or surrogate, $P \in \R^k$ be the observed
k-dimensional vector of noisy proxies, and Y be the scalar long-term
outcome. Potential outcomes are indexed $Z(w)$ for $Z \in \{S, P, Y\}$ and $
w \in \{0,1\}$. We assume $W$ is independent from all potential outcomes due to random treatment assignment.

The core relationships we assume are:
\begin{enumerate}
\item \textbf{First Stage (Treatment $\rightarrow$ Surrogate)}: The
treatment $W$ has a linear causal effect on the true surrogate: $S(w) =
\gamma w + \epsilon_S$, where $\gamma \in \R^p$ is the non-random vector of
treatment effects on the latent surrogates and $\epsilon_S$ is a random residual vector.
\item  \textbf{Proxy Generation (Surrogate $\rightarrow$ Proxies)}:
The $k$ observed proxies are noisy linear projections of the true
surrogate: $P(w) = LS(w) + E$, where $L \in \R^{k\times p}$ is a non-random loading
matrix and E is a random $k$-dimensional noise residual. We focus on the high-dimensional
case where the number of proxies is large---in particular, it is larger than the number of surrogates, $k>p$.
\item  \textbf{Second Stage (Surrogate $\rightarrow$ Outcome)}: The
true surrogate has linear causal effects on the long-term outcome: $Y(w)
= S(w)^\tra \beta + \eta$, where non-random $\beta \in \R^p$ defines the ``true
surrogate index,'' $S^\tra \beta$ and $\eta$ is a random residual.

\end{enumerate}

Note that the model imposes a conventional surrogacy condition: that $W$ only affects $Y$ through $S$. Thus, under the model, the long-term treatment effect is the total effect
of $W$ on $Y$ as mediated by $S$:
\begin{equation*}
 \tau^{*} = Y(1)-Y(0)=\gamma^\tra \beta.
\end{equation*}

\paragraph{Economic Foundations.} The linear structure above admits a natural interpretation through a
random utility framework \parencite{mcfadden1974conditional}. Consider
a consumer $i$ whose preferences over a firm's offerings are
summarized by a $p$-dimensional \emph{preference state} $S_i$. Each
dimension of $S_i$ captures a distinct aspect of the consumer's latent
valuation---for instance, perceived product quality, price
sensitivity, or brand affinity. The treatment $W_i$ (e.g., exposure to a new recommendation algorithm,
a promotional offer, or an improved user interface) shifts the
consumer's preference state:
\begin{equation*}
    S_i = \bar{S} + \gamma W_i + \tilde\epsilon_{S,i}
\end{equation*}\vspace{-0.7cm}

where $\bar{S}$ is the baseline preference state and $\gamma \in \R^p$
captures how the treatment moves each preference dimension.\footnote{This
formulation nests the standard case where treatment affects a single
index of ``customer satisfaction'' ($p=1$) as well as richer settings
where treatment differentially affects multiple preference dimensions.} The firm observes a high-dimensional vector of short-term behavioral
outcomes $P_i \in \R^k$: e.g., clicks, page views, add-to-cart events,
purchases across categories, or session durations. These
behaviors are \emph{choice outcomes} that reflect the consumer's
underlying preferences. Under a linear approximation to the demand
system---valid either as a first-order expansion around the mean or
under quasi-linear utility with linear demand
\parencite{berry1994estimating}---observed behaviors load on the
latent preference state:
\begin{equation*}
    P_i = L S_i + E_i
\end{equation*}\vspace{-0.7cm}

The loading matrix $L \in \R^{k \times p}$ captures how each
preference dimension manifests in observable behavior. For example, if
$S_{i1}$ represents price sensitivity, the corresponding column of $L$
would place high weight on behaviors related to deal-seeking (coupon
usage, price comparison clicks) and low weight on premium purchases.
The noise term $E_i$ reflects idiosyncratic variation in behavior not
explained by preferences---measurement error, contextual factors, or
random utility shocks at the choice level.\footnote{This factor structure is standard in psychometric and econometric
settings where many noisy measurements proxy for fewer latent
constructs \parencite{bai2002determining, anderson2003introduction}.} Finally, the firm's long-term outcome of interest, $Y_i$ (e.g., customer
lifetime value, retention, or future spending), aggregates the value
generated by a consumer with preference state $S_i$:
\begin{equation*}
    Y_i = S_i^\tra \beta + \eta_i
\end{equation*}\vspace{-0.7cm}

The coefficient vector $\beta$ captures how each preference dimension
contributes to long-term value: e.g., consumers with higher quality
perceptions or lower price sensitivity may generate more lifetime
value.

\paragraph{The Identification Challenge.}
The economic structure clarifies why the surrogate inference problem
is non-trivial. The firm wishes to estimate $\tau^* = \gamma^\tra
\beta$, the causal effect of treatment on long-term value. However,
the preference state $S_i$ is unobserved and  the structural parameters
$(\gamma, \beta)$ are unknown. The firm can only estimate
reduced-form objects: treatment effects on proxies ($\tau_P =
L\gamma$) and the predictive relationship between proxies and outcomes
($\alpha$, from regressing $Y$ on $P$). The question is whether these
reduced-form quantities can be combined to recover the structural
effect $\tau^*$ and, if so, how measurement error in the proxies
affects the resulting estimator.

\subsection {Data Structure and the Feasible Estimator}

We assume we have access to two
datasets with observations $i$ drawn \emph{i.i.d.} from the common model:
\begin{itemize}
    \item \textbf{Experimental Dataset} ($\mathcal{D}_E$): A sample of
    $N_e$ units with observations ($W_i, P_i$). This lets us
    estimate the first-stage effect of the treatment on the observable
    proxies, $\tau_p = \E[P|W=1] - \E[P|W=0]$. 
    \item \textbf{Observational Dataset} ($\mathcal{D}_o$): A
    separate, larger sample of $N_o>N_e$ units with observations ($P_i,
    Y_i$). This lets us estimate the relationship between the
    proxies and the long-term outcome.
\end{itemize}

Specifically, using the latter dataset, we can obtain regularized estimates of the regression of $Y$ on $P$ (recalling $P$ is high-dimensional). For example, we can estimate a ridge regression with regularized coefficient vector:
\begin{equation}
\hat{\alpha}_\lambda = \left(\sum_i P_i P_i^\tra + \lambda I_k\right)^{-1} \sum_i P_i Y_i,
\label{eqn:surrind}
\end{equation}
for some tuning parameter $\lambda$.

Our feasible estimator for the long-term treatment effect combines
information from both stages. We first estimate the feasible surrogate
index by regressing Y on P to get coefficients  as in equation
\eqref{eqn:surrind}. Then, we estimate the treatment effect on the
proxies, $\hat{\tau}_P$, from the experimental data. The final
estimator is:
$$
\hat{\tau} = \hat{\tau}_p^\tra \hat{\alpha}_\lambda.
$$
where $\hat{\tau}_p$ gives the difference in sample means of $P_i$ for $W_i=1$ vs $W_i=0$ in the experimental dataset.

It is straightforward to show that such an estimator can identify the true long-run effect $\tau^*$ in the absence of proxy measurement error. Specifically, when $E=0$ the proxies $P=LS$ satisfy the usual surrogacy condition and a non-regularized regression of $Y_i$ on $P_i$ identifies $(L^\prime)^+\beta$ while $\hat\tau_P$ identifies $L\gamma$.\footnote{Here $V^+$ denotes the Moore-Penrose pseudoinverse of $V$ and we assume $L$ has full column rank.} Hence we identify:
\begin{align*}
(L\gamma)^\prime (L^\prime)^+\beta=\gamma^\prime\beta=\tau^*.
\end{align*}
Intuitively, with perfect
proxies, the surrogate index approach recovers the true long-term
effect. In practice, however, measurement error in the proxies make the feasible
estimator biased. This leads to the central questions of this
paper:
\begin{enumerate}
    \item \textbf{Bias characterization}: How large is the attenuation
    bias when proxies are noisy?
    \item \textbf{The role of many proxies}: Does having more proxies
    $k$ reduce this bias?
    \item \textbf{Estimation}: How should we estimate $\alpha$ when
    $P$ is high-dimensional and noisy?
\end{enumerate}

We address these questions in Section~\ref{sec:theory}. We first build intuition with a scalar surrogate model, showing that attenuation bias shrinks as more proxies are added. We then provide our main theoretical result: a closed-form characterization of the bias-variance tradeoff for Ridge regression. Section~\ref{sec:simulations} validates these findings through numerical experiments, and Section~\ref{sec:empirical} applies our methods to the California GAIN experiment.

\FloatBarrier

\section{Theoretical Analysis}\label{sec:theory}

\subsection{Motivating Cases}\label{sec:warmup}

To build intuition for the high-dimensional setting, we first analyze
a simple case with a scalar surrogate ($p=1$) and two noisy
proxies ($k=2$). We aim to understand how a non-regularized estimator combines these
proxies to approximate the true surrogate S. The insights from this
simple case highlight the inherent challenges from measurement error and
motivate the need for a more systematic approach, with regularization, for large $k$.

In the simple model, the treatment has a causal effect of $\tau_S$ on the true
surrogate and we observe noisy versions of this surrogate:
\begin{align*}
    S & = \tau_S W + \varepsilon_S \nonumber \\
    P_1 & = S + \varepsilon_1 \\
    P_2 & = S + \varepsilon_2
\end{align*}
The
errors $\varepsilon_1, \varepsilon_2$ are uncorrelated with each
other, with the true signal $S$, and with the outcome error $\eta$.
Their means are zero and the variances are $\Var(S)=\sigma_S^2$,
$\Var(\varepsilon_1)=\sigma_1^2$, $\Var(\varepsilon_2)=\sigma_2^2$.
The true and feasible surrogate index in this setting are:
\begin{align*}
    Y & = \beta S + \eta  \\ Y & = \alpha_1
    \underbrace{P_1}_{S + \varepsilon_1} + \alpha_2
    \underbrace{P_2}_{S+ \varepsilon_2} + \eta
    \label{eqn:emp_surind}
\end{align*}

We seek to approximate the true surrogate index $\beta S$
with the feasible surrogate index $\alpha_1 P_1+\alpha_2 P_2$, which is an
errors-in-variables model with measurement error in the regressors \parencite{fuller1987measurement}. 

\paragraph*{Deriving the Coefficient Magnitudes}

The feasible regression coefficients $(\alpha_1, \alpha_2)$ are biased relative to the true effect $\beta$, due to the measurement error in $P_1$ and $P_2$. Specifically, Appendix
\ref{app:alpha_solve} shows 
\begin{align*}
    \alpha_1 &= \beta \frac{\sigma_S^2
    \sigma_2^2}{\sigma_S^2\sigma_1^2 + \sigma_S^2\sigma_2^2 +
    \sigma_1^2\sigma_2^2} \\
    \alpha_2 &= \beta \frac{\sigma_S^2
    \sigma_1^2}{\sigma_S^2\sigma_1^2 + \sigma_S^2\sigma_2^2 +
    \sigma_1^2\sigma_2^2},
\end{align*}
implying the following:
\begin{enumerate}
   \item \textbf{Relative Magnitudes} 
    \begin{equation*}
        \frac{\alpha_1}{\alpha_2} = \frac{\sigma_2^2}{\sigma_1^2}
    \end{equation*}
    The relative size of the estimated coefficients is the inverse
    ratio of their measurement error variances. The proxy with less
    noise (smaller error variance) gets the larger coefficient. 

    \item \textbf{Aggregate Attenuation} The sum of the coefficients
    is still less than the true effect $\beta$:
    \begin{equation*}
        \alpha_1+\alpha_2 = \beta
        \frac{\sigma_S^2(\sigma_1^2 +
        \sigma_2^2)}{\sigma_S^2(\sigma_1^2 + \sigma_2^2) +
        \sigma_1^2\sigma_2^2} < \beta
    \end{equation*} The total effect is attenuated (biased
    towards zero).
\end{enumerate}



It
follows directly that the feasible estimator for the total effect will be inconsistent, with attenuation bias:
\begin{equation*}
    \plim(\tauhat_p^\prime \hat\alpha) = \tau_S(\alpha_1+\alpha_2) <
    \tau_S \beta = \tau^*
\end{equation*}

This simple example yields two insights. First, the estimated long-run effect is generally attenuated
relative to the true $\beta$, a classic consequence of measurement error. Second, the proxy regression  assigns a larger coefficient to the proxy with lower measurement error variance. This provides a theoretical basis for the intuition that one should upweight more reliable proxies. We now generalize this scalar surrogate model to show how increasing the number of proxies, $k$, while keeping the true surrogate as a scalar, to study how many noisy proxies help with the attenuation.

\subsubsection*{Scaling to Many Proxies: Attenuation Bias Shrinks with
$k$}

We now maintain the scalar surrogate model ($p=1$) but expand to $k$ noisy
proxies:
\begin{align*} S &= \tau_S W + \varepsilon_S, \quad \text{with }
    \mathrm{Var}(S) = \sigma_S^2 \\ P_j &= S + \varepsilon_j, \quad
    \text{for } j=1, \dots, k \\ Y &= \beta S + \eta
\end{align*}
To facilitate a clear derivation, we assume the proxy errors
$\varepsilon_j$ are independent and identically distributed with
$\mathrm{Var}(\varepsilon_j) = \sigma_E^2$. Let the vector of proxies
be $P \in \mathbb{R}^k$. The population regression coefficients for a
regression of $Y$ on $P$ are given by $\alpha = [\mathrm{Var}(P)]^{-1}
\mathrm{Cov}(P, Y)$. First, we find the necessary variance and
covariance terms:
\begin{itemize}
    \item The covariance of each proxy with the outcome is identical:
    $\mathrm{Cov}(P_j, Y) = \mathrm{Cov}(S + \varepsilon_j, \beta S +
    \eta) = \beta\sigma_S^2$. Thus, the covariance vector is
    $\mathrm{Cov}(P, Y) = \beta\sigma_S^2 \mathbf{1}_k$, where
    $\mathbf{1}_k$ is a $k \times 1$ vector of ones.
    \item The $k \times k$ variance-covariance matrix of the proxies
    $P$ has diagonal elements $\mathrm{Var}(P_j) = \sigma_S^2 +
    \sigma_E^2$ and off-diagonal elements $\mathrm{Cov}(P_i, P_j) =
    \sigma_S^2$. This patterned matrix can be written compactly as
    $\mathrm{Var}(P) = \sigma_S^2 J_k + \sigma_E^2 I_k$, where $J_k$
    is the $k \times k$ matrix of ones and $I_k$ is the identity
    matrix.
\end{itemize}
Using the Sherman-Morrison formula, the inverse of this matrix is:
\begin{equation*}
    [\mathrm{Var}(P)]^{-1} = \frac{1}{\sigma_E^2}I_k - \frac{\sigma_S^2}{\sigma_E^2(\sigma_E^2 + k\sigma_S^2)}J_k
\end{equation*}

Solving for the coefficient vector $\alpha$ yields:
\begin{align*}
    \alpha &= \left( \frac{1}{\sigma_E^2}I_k -
           \frac{\sigma_S^2}{\sigma_E^2(\sigma_E^2 + k\sigma_S^2)}J_k
           \right) (\beta\sigma_S^2 \mathbf{1}_k) \nonumber \\ &=
           \left( \frac{\beta\sigma_S^2}{\sigma_E^2} -
           \frac{\beta\sigma_S^4(k)}{\sigma_E^2(\sigma_E^2 +
           k\sigma_S^2)} \right) \mathbf{1}_k \nonumber \\ &=
           \frac{\beta\sigma_S^2}{\sigma_E^2 + k\sigma_S^2}
           \mathbf{1}_k
\end{align*}
This shows that regression correctly assigns an equal coefficient
$\alpha_j$ to every proxy due to the symmetry of the problem. The sum
of these coefficients, representing the aggregate effect of the latent
surrogate on the outcome, is:
\begin{equation*}
    \sum_{j=1}^k \alpha_j = k \cdot \left(
    \frac{\beta\sigma_S^2}{\sigma_E^2 + k\sigma_S^2} \right) = \beta
    \left( \frac{k\sigma_S^2}{k\sigma_S^2 + \sigma_E^2} \right)
\end{equation*}

This expression reveals a central insight: the total effect is
attenuated by a factor that depends on the signal-to-noise ratio and,
crucially, on the number of proxies $k$. As $k \to \infty$, this 
converges to the true $\beta$:
\begin{equation*}
    \lim_{k \to \infty} \sum_{j=1}^k \alpha_j = \beta \cdot \lim_{k
    \to \infty} \frac{k\sigma_S^2}{k\sigma_S^2 + \sigma_E^2} = \beta
\end{equation*}
This derivation demonstrates that as the number of noisy proxies increases, ordinary least squares (OLS)
can average out the measurement error and recover the true underlying
relationship. This is the core intuition behind our general result in
the next section: a larger number of proxies per surrogate (a larger
$c$) reduces the attenuation bias. For any finite $k$, however, some
bias remains, and the multicollinearity of the proxies motivates the
use of regularized methods to produce a stable surrogate index.


\subsection{High-Dimensional Selection with the Ridge Surrogate Index}\label{sec:ridge}

The 2-proxy case provides the intuition that coefficient magnitudes
are inversely related to noise. We now generalize this to a
high-dimensional setting to see how regularization can help
manage this trade-off. Specifically, we consider ridge regression and correspondingly adopt normality assumptions to ease derivations. Here we allow for multiple surrogates.






\subsubsection{Model and Estimator}
Consider again a setting with a $p$-dimensional unobserved surrogate $S$, a $k$-dimensional vector of observed proxies $P$, a binary treatment $W$, and a scalar outcome $Y$. The \emph{i.i.d.} data generating process  is as follows:
\begin{enumerate}
    \item \textbf{Surrogate Generation}: $S_i = W_i^\prime\gamma + \epsilon_S$ where $\epsilon_S \mid W_i \sim \mathcal{N}(0, \Sigma_S)$. Here $\gamma \in \R^p$ is the vector of treatment effects on the surrogates. 
    \item \textbf{True Surrogate Index}: $Y_i = S_i^\tra \beta + \eta_i$, where $\beta \in \R^p$ and $\eta_i\mid S_i,W_i \sim \mathcal{N}(0, \sigma_\eta^2)$. The long-run treatment effect is $\tau^* = \gamma^\tra \beta$.
    \item \textbf{Proxy Model}: $P_i = L S_i + E_i$, where $L \in \R^{k \times p}$ is a loading matrix and $E_i \mid W_i,S_i ,\eta_i \sim \mathcal{N}(0, \Sigma_E)$.
\end{enumerate}

Our feasible long-term treatment effect estimator combines a
first-stage estimate of the effect of $W$ on each proxy,
$\hat{\tau}_j$, with a second-stage Ridge regression of $Y$ on $P$:
$$
\hat{\tau}_\lambda =
\hat{\tau}_p^\tra \hat{\alpha}_\lambda
=  \sum_{j=1}^k \hat{\tau}_j \hat{\alpha}_{\lambda, j} \quad \text{where} \quad \hat{\alpha}_\lambda = \left(\sum_i P_i P_i^\tra + \lambda I_k\right)^{-1} \sum_i P_i Y_i
$$

\subsubsection{Asymptotic Bias}
As $N_0 \to \infty$, the estimator converges to its population analogue, $\tau_\lambda = \tau_P^\tra \alpha_\lambda$, where $\tau_P = L\gamma$ and $\alpha_\lambda = (\Sigma_{PP} + \lambda I_k)^{-1} \Sigma_{PY}$. The general form of the asymptotic bias, $\text{Bias}_\lambda = \tau_\lambda - \tau^*$, is:
\begin{equation*}
    \text{Bias}_\lambda = \gamma^\tra  \left[ L^\tra (L \Sigma_S L^\tra + \Sigma_E + \lambda I_k)^{-1} L \Sigma_S - I_p \right] \beta \label{eqn:bias_gen}. 
\end{equation*}

While general, this expression is difficult to interpret.  The bias
has a complex general form that depends on the loading structure,
noise covariance, and regularization. By making simplifying
assumptions of homoskedasticity ($\Sigma_S = \sigma^2_S I_p$,
$\Sigma_E = \sigma^2_E I_k$), we can derive a simple analytic
expression for bias.

\begin{remark}{Spectral Bias Decomposition}

Let $L = U D V'$ be the SVD with singular values $d_1, \dots, d_p > 0$. Under homoskedasticity, we can define rotated coordinates $\tilde{\gamma} = V'\gamma, \tilde{\beta} = V'\beta$. Then,
$$
\tau_\lambda = \sum_{j=1}^p \frac{\sigma_S^2 d_j^2}{\sigma_S^2 d_j^2 +
\sigma_E^2 + \lambda} \tilde{\gamma}_j \tilde{\beta}_j
$$
and the corresponding bias is
$$
\text{Bias}_\lambda = - \sum_{j=1}^p (1-\kappa_j(\lambda))  \tilde{\gamma}_j \tilde{\beta}_j
$$

Interpretation:
\begin{itemize}
\item Each latent direction $j$ has its own shrinkage factor
$\kappa_j(\lambda)$; directions with small $d_j^2$ (that are weakly
measured by proxies) are attenuated more.
\item Adding proxies that increase the smallest
$d_j^2$ (better coverage of the true $S$) help the most; adding
redundant proxies that only inflate already large singular values do
not.
\end{itemize}
\end{remark}

Finally, if we make an additional assumption of balanced loadings
($L^\tra L = c I_p$)\footnote{this requires that the proxies are
conditionally orthogonal after accounting for the latent surrogates},
the bias reduces to a remarkably simple form (see Appendix
\ref{app:bias_derivation} for the full derivation):
\begin{equation}
    \text{Bias}_\lambda = - \left( \frac{\sigma_E^2 + \lambda}{c
    \sigma_S^2 + \sigma_E^2 + \lambda} \right) \tau^{*}
    \label{eqn:bias_lam}
\end{equation}
This result reveals several key properties of the estimator:
\begin{itemize}
    \item The estimator is always attenuated towards zero.
    \item The bias is directly proportional to the true treatment
    effect, $\tau^*$.
    \item Larger $c$ (more proxies per true unobserved surrogate) reduces bias by
    strengthening the signal. As we let $c \to \infty$, this bias
    disappears
    \item Regularization $\lambda > 0$ increases the magnitude of the
    attenuation bias.
\end{itemize}

\subsubsection{Asymptotic Mean Squared Error (AMSE)} Our primary goal
is to minimize the total error of the estimator. By deriving the
asymptotic variance and combining it with the squared bias, we arrive
at the full Asymptotic Mean Squared Error (AMSE), as detailed in
Appendix \ref{app:asym_deriv}.
\begin{equation*}
\boxed{
    \AMSE(\hat{\tau}_{\lambda}) = \underbrace{ \left( \frac{\sigma_E^2
+ \lambda}{c\sigma_S^2 + \sigma_E^2 + \lambda} \right)^2 (\tau^*)^2
}_{\text{Squared Bias}} + \underbrace{ \frac{\sigma_\eta^2 c
||\vect{\gamma}||^2}{n} \frac{c\sigma_S^2 + \sigma_E^2}{(c\sigma_S^2 +
\sigma_E^2 + \lambda)^2} }_{\text{Variance}} }
\end{equation*}

This expression formalizes the classic bias-variance trade-off. The
bias term is monotonically increasing in $\lambda$, while the variance
term is monotonically decreasing in $\lambda$. Minimizing this AMSE
with respect to $\lambda$ yields the optimal regularization parameter
(see Appendix \ref{app:lambda_derivation} for the derivation):
\begin{equation*}
\boxed{
    \lambda^* = \frac{\sigma_\eta^2 ||\vect{\gamma}||^2 (c\sigma_s^2 + \sigma_e^2)}{n (\tau^*)^2 \sigma_s^2} - \sigma_e^2
}
\end{equation*}

Since $\lambda^*$ depends on unobservable population quantities, this
theoretical result provides the justification for using data-driven
methods like cross-validation to select $\lambda$ in practice.

\subsection{Variance estimation and inference}\label{sec:variance}

Next, we propose a simple large-sample variance estimator for our
two-sample ridge estimator
$$
\hat\tau_\lambda \;\equiv\; \hat\tau_P^\top \hat\alpha_\lambda,
$$

where $\hat\tau_P$ is the estimated treatment effect on the proxies in
the experimental sample $D_E$ and $\hat\alpha_\lambda$ is the ridge
coefficient vector from the observational sample $D_O$. Throughout, we
assume $D_E$ and $D_O$ are independent (no overlapping units). We
target inference for the population estimand $\tau_\lambda =
\tau_P^\top \alpha_\lambda$, i.e., the estimand delivered by the ridge
procedure.

\paragraph{Delta-method decomposition.}

Treating $\hat\tau_\lambda$ as a smooth function of
$(\hat\tau_P,\hat\alpha_\lambda)$, independence implies
$$
\Var(\hat\tau_\lambda)\;\approx\; \Var\!\big(\hat\tau_P^\top
\alpha_\lambda\big) + \Var\!\big(\tau_P^\top \hat\alpha_\lambda\big)
\;\approx\; \alpha_\lambda^\top \Var(\hat\tau_P)\alpha_\lambda \;+\;
\tau_P^\top \Var(\hat\alpha_\lambda)\tau_P.
$$

We estimate this by a plug-in estimator:
\begin{equation}\label{eq:var_decomp}
\widehat{\Var}(\hat\tau_\lambda)\;\equiv\; \hat\alpha_\lambda^\top
\widehat{\Var}(\hat\tau_P)\hat\alpha_\lambda \;+\; \hat\tau_P^\top
\widehat{\Var}(\hat\alpha_\lambda)\hat\tau_P.
\end{equation}

\paragraph{First-stage variance (experimental sample).}

In the simplest difference-in-means implementation,
$\hat\tau_P = \bar P_1 - \bar P_0$ (componentwise), where $\bar P_w$ is the sample mean of $P$ among units with $W=w$. Then
\begin{equation}\label{eq:var_tauP}
\widehat{\Var}(\hat\tau_P)\;\equiv\; \frac{\widehat{\Sigma}_{P\mid W=1}}{n_1} + \frac{\widehat{\Sigma}_{P\mid W=0}}{n_0},
\end{equation}

with $n_w$ the number of treated/control units and
$\widehat{\Sigma}_{P\mid W=w}$ the sample covariance of $P$ within arm
$w$. More generally, if $\hat\tau_P$ is obtained by a (columnwise)
regression of $P$ on $(1,W)$ (or on $(1,W,X)$ with covariates),
$\widehat{\Var}(\hat\tau_P)$ is the usual sandwich covariance for the
coefficient on $W$.

\paragraph{Second-stage variance (observational ridge).}

Let $X$ denote the $n_O\times k$ design matrix of proxies in $D_O$ and
$y$ the $n_O$-vector of outcomes. The ridge estimator is
$\hat\alpha_\lambda=(X^\top X+\lambda I)^{-1}X^\top y$. Write
$A_\lambda\equiv (X^\top X+\lambda I)^{-1}$ and residuals $\hat e
\equiv y - X\hat\alpha_\lambda$. A heteroskedasticity-robust
(Eicker--Huber--White) variance estimator for $\hat\alpha_\lambda$ is
\begin{equation}\label{eq:var_alpha}
\hat{\Var}(\hat{\alpha}_\lambda)\;\equiv\; A_\lambda (X^\top
\text{diag}(\hat{e}_1^2,\ldots,\hat{e}_{n_O}^2) \; X )A_\lambda
\end{equation}

Under homoskedasticity, one may replace $X^\top
\text{diag}(\hat{e}^2)X$ with $\hat\sigma^2 X^\top X$, where
$\hat\sigma^2 = n_O^{-1}\sum_{i=1}^{n_O}\hat e_i^2$.

\paragraph{Standard errors and confidence intervals.}

Combining \eqref{eq:var_decomp}--\eqref{eq:var_alpha} yields a
feasible standard error $\hat{\text{SE}}(\hat\tau_\lambda) \equiv
\sqrt{\hat{\Var}(\hat\tau_\lambda)}$. A Wald interval is
$\hat\tau_\lambda \pm z_{1-\alpha/2}\hat{\text{SE}}(\hat\tau_\lambda)$.
When $D_E$ and $D_O$ overlap, an additional covariance term arises in
\eqref{eq:var_decomp}; in that case, sample splitting (or explicitly
estimating the overlap covariance) restores validity.





\section{Numerical Experiments}\label{sec:simulations}

\subsection{MSE for Treatment Effect Parameter}

We have a randomly assigned binary treatment $W$, which has a
treatment effect $\mathbf{\tau}_S$ on the true surrogate. The
long-term outcome $Y$ is generated  from the true surrogate index $Y_i
= S_i^\tra \beta + \varepsilon_i$. We observe a large vector $P$
generated from the a low-rank representation $S \cdot L^\tra + E$
where $L$ is a set of random loadings and $E$ is multivariate normal
noise with some covariance $\Sigma_E$. We first estimate treatment
effects on each proxy via OLS $P_j \sim \tau_j W$, then estimate a
feasible surrogate index model $Y \sim P \alpha + \epsilon$. Finally,
we combine the first-stage and second stage models to estimate the
long-term effect:
\begin{equation}
 \hat{\tau}^{\text{Longterm}} = \sum_{j \in \mathcal{J}} \tau_j \alpha_j   \label{eqn:lt_surro}
\end{equation}

where $\alpha_j$ is the estimated surrogate index coefficient
for proxy $P_j$. We begin with the following parameters:
\begin{itemize}
    \item n = 5000 observations
    \item p = 5 true surrogates (parametrized)
    \item k = 30 observed proxies (6 proxies for each surrogate, parametrized)
    \item $W$ is a binary treatment
    \item $S ~ N(0, \Sigma_S) + \tau W$ with random Toeplitz covariance $\Sigma_S$, with treatment effect $\tau_S = 0.5$.
    \item $\beta = [1, 1/2, 1/3, 1/4, 1/5]'$ (decaying importance)
    \item $\varepsilon \sim N(0, 0.5^2)$
    \item Proxy : $P = S L + E$ where
    \begin{itemize}
        \item  $L: k\times p$ loading matrix (block structure + noise)
        \item $E ~ N(0, \Sigma_E)$ with compound symmetry: 
            $\Sigma_E[i,j] = \sigma^2 (1 + \rho 1[i\neq j])$
        \item $\text{SNR} = \Var(SL)/\Var(E)$ is the signal-to-noise ratio.
    \end{itemize}
\end{itemize}

These numerical experiments test the methods under more general
conditions (e.g., correlated proxy errors) than the specific
simplifying assumptions used in the theoretical derivations for Ridge
regression, and

We compare the following methods used to estimate the feasible
surrogate model $Y \sim \alpha P$.

\begin{enumerate}

    \item \textit{Oracle Direct Regression}: $Y \sim W$, which
    directly gives us the treatment effect but is infeasible since we
    don't observe $Y$ and $W$ in the same dataset. This is reported as
    the blue line in all simulation figures.
    
    \item \textit{OLS Screening/L0 Regression}: We select the top 10
    proxies by $\mid \alpha_j \mid$ as the selected proxies
    $\mathcal{J}$, and then compute the treatment effect with
    \eqref{eqn:lt_surro}
    \item \textit{Lasso/L1 Regression}: We regress the long-term
    outcome $Y$ on all proxies $P$ with cross-validated LASSO
    regression (L1), so the selected model $\mathcal{J}$ is sparse,
    and then compute the treatment effect with \eqref{eqn:lt_surro}.
    \item \textit{Ridge/L2 Regression}: We regress the long-term
    outcome $Y$ on all proxies $P$ with cross-validated ridge
    regression (L2), so the selected model $\mathcal{J}$ is the entire
    set of proxies, and then compute the treatment effect with
    \eqref{eqn:lt_surro}.
    
    \item \textit{Partial Least Squares}:  Partial Least Squares
    (PLS): This method reduces the dimensionality of the proxies P by
    constructing a set of latent components that maximize the
    covariance with the outcome Y. We then regress Y on these
    components and use the resulting model for the second stage
    \parencite{hastie2009elements}. This gives us a modified version
    of principal-components regression that uses covariance
    information with the target $y$ rather than the covariance matrix
    of $X$ alone. We then plug these into \eqref{eqn:lt_surro}
\end{enumerate}

\begin{figure}
    \centering
    \includegraphics[width=\linewidth]{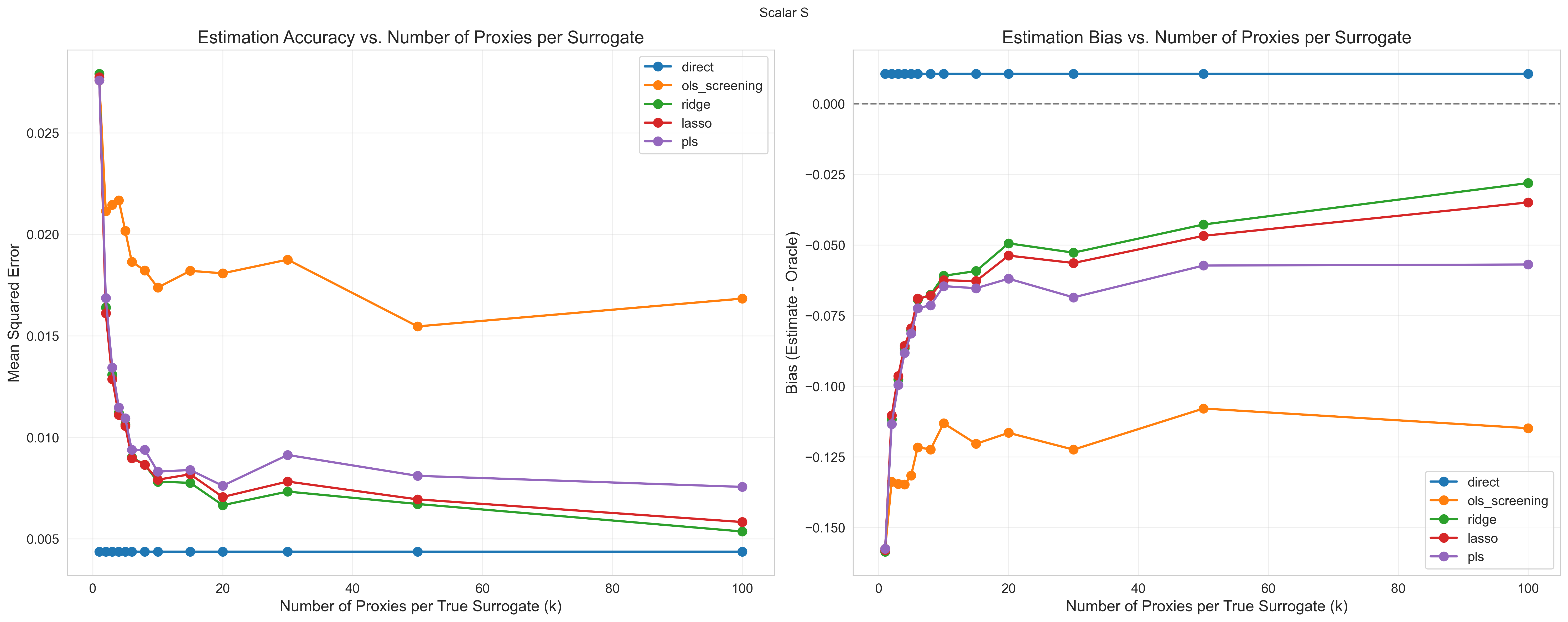}
    \caption{Estimation Performance when increasing number of proxies for 1 scalar surrogate $S$}
    \label{fig:te_varyk_fixp1}
\end{figure}

\begin{figure}
    \centering
    \includegraphics[width=\linewidth]{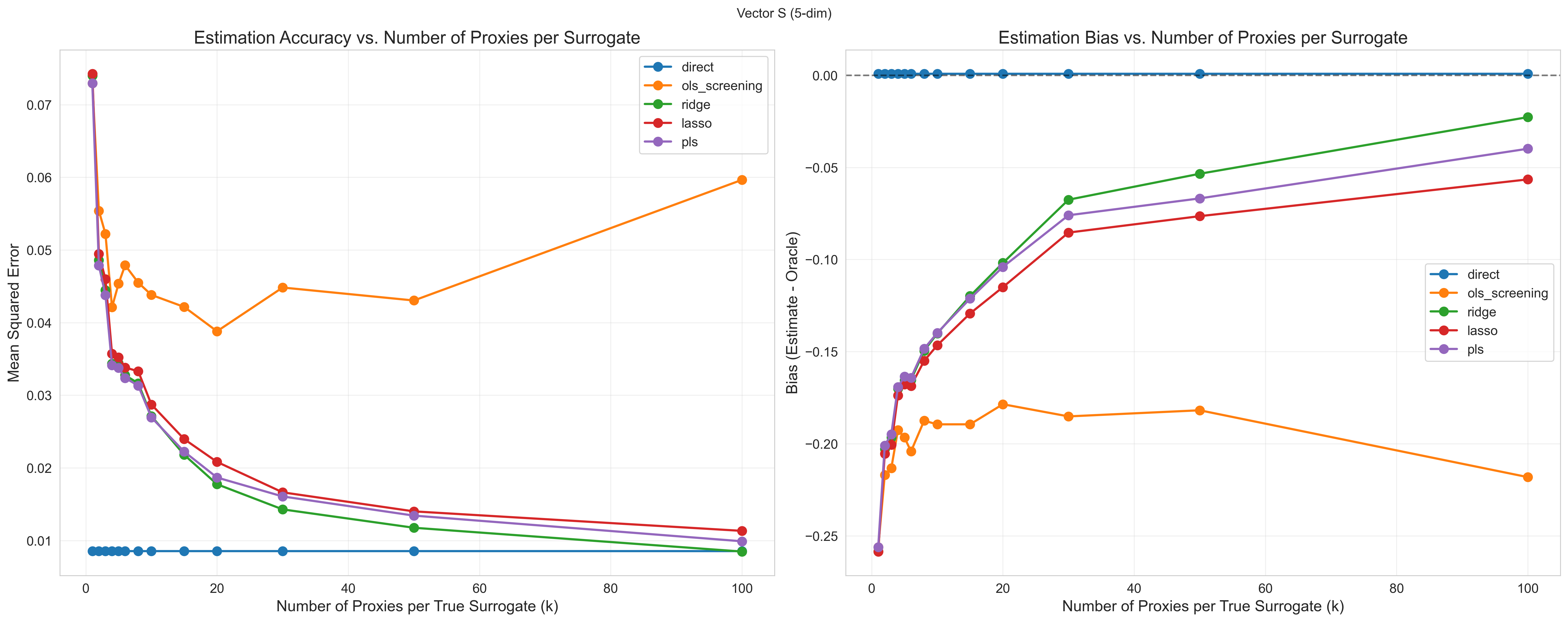}
    \caption{Estimation Performance when increasing number of proxies for a 5-vector-valued surrogate $S$}
    \label{fig:te_varyk_fixp5}
\end{figure}

\begin{figure}
    \centering
    \includegraphics[width=\linewidth]{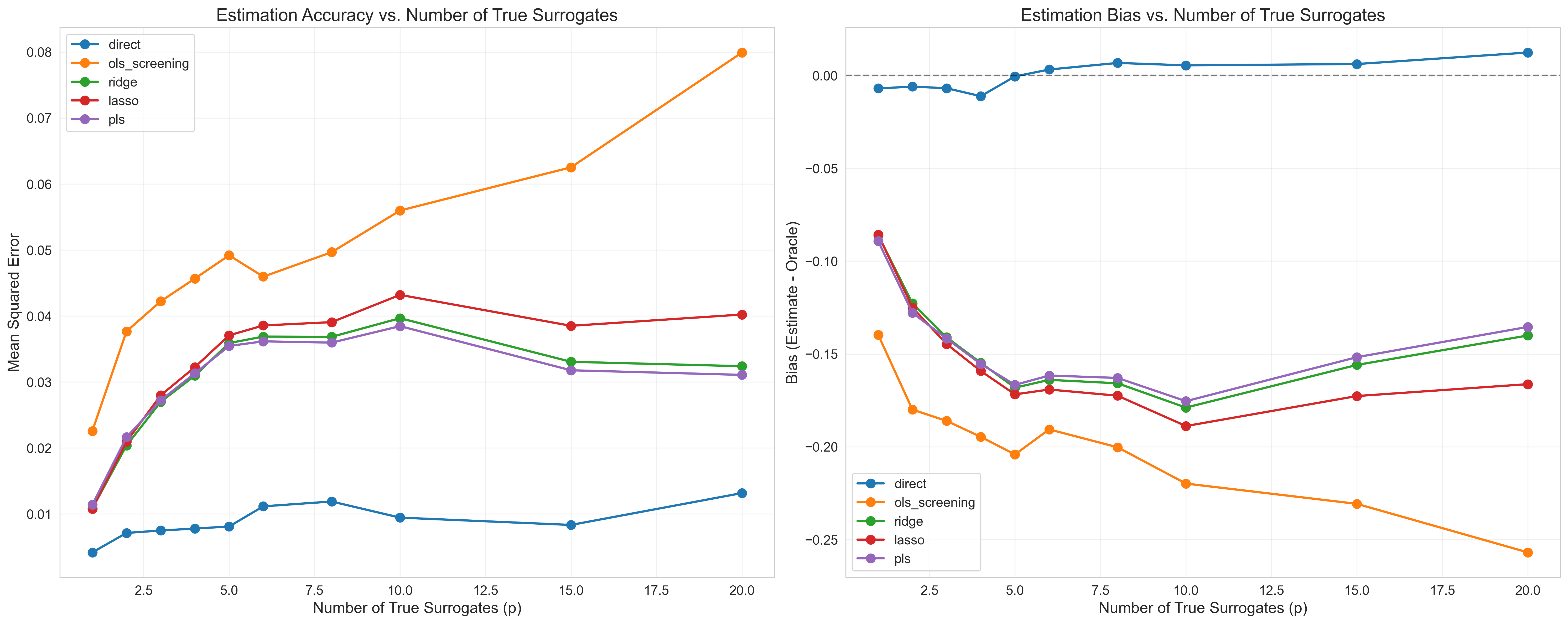}
    \caption{Estimation Performance when increasing number of surrogates while fixing the number of proxies for each surrogate at 6}
    \label{fig:te_fixk_varyp}
\end{figure}

The simplest case where we study the RMSE relative to the true
treatment effect where we have a scalar-valued surrogate and vary the
number of proxies $P$. We report bias and RMSE in
Figure~\ref{fig:te_varyk_fixp1}.  The oracle estimator estimates the
true long-term treatment effect, and we evaluate the different
feasible estimators based on how closely they approach the oracle as
the number of proxies increases. Ridge and Partial-Least-Squares are
best performing. They slightly outperform both Lasso (L1) and OLS
(L0), where model selection mistakes in the surrogate index introduce
bias into the treatment effect estimate. In general, however, there is
little separation between the methods; they all perform better and get
close to the oracle in RMSE terms as we increase the number of
observed proxies.

This empirical result directly confirms our theoretical result in
equation \eqref{eqn:bias_lam}. The bias term, proportional to $(\sigma_e² +
\lambda) / (c \sigma^2 + \sigma_e^2 + \lambda)$, clearly shows that as
the number of proxies per surrogate increases, the `balanced loadings'
constant c also increases, causing the asymptotic bias to shrink
towards zero. The regularized estimators successfully leverage this by
averaging across many noisy signals.

Conversely, Figure~\ref{fig:te_varyk_fixp5} highlights the fundamental
weakness of OLS screening. As k increases, the total number of
available proxies (k$\times$p) grows, increasing the likelihood of
spurious correlations between some proxies and the outcome Y. OLS, by
greedily selecting proxies based on the largest (and potentially
spurious) coefficients, is prone to model selection errors. This
instability leads to a deterioration in performance, a failure mode
that regularized methods are specifically designed to prevent.

When we have a vector-valued true surrogate as in
Figure~\ref{fig:te_varyk_fixp5}, we have more separation between the
methods: ridge and PLS are best-performing, closely followed by LASSO.
OLS screening is worst-performing in this case and deteriorates when
we increase the number of proxies per surrogate. In the
increasing-numbers-of-proxies regime, we find that feasible surrogate
indices (Ridge/PLS/Lasso but not OLS screening) approach the MSE of
the oracle estimator.

Figure~\ref{fig:te_fixk_varyp} illustrates a more challenging scaling
problem that shows the limits of our approach. As the dimension of the
true surrogate p increases with a fixed number of proxies per
surrogate (6 proxies per surrogate), the complexity of the true signal
that must be recovered increases. While the total number of proxies
also grows, the amount of information per true surrogate dimension
remains constant. This makes it fundamentally harder for the models to
disentangle the p different signal channels, leading to slower
convergence and persistent bias, even for the best-performing methods.
Nevertheless, the rank ordering of methods from the previous setting
persists: ridge regression still has the lowest RMSE.

\FloatBarrier

Finally, we jointly vary the number of proxies (x-axis) and true
surrogates (y-axis) in
Figures~\ref{fig:heatmap_ols}--\ref{fig:heatmap_pls} in
Appendix~\ref{appdx:gridsim}. These heatmaps provide a comprehensive
view of estimator performance. In Figure~\ref{fig:heatmap_ridge} for
Ridge, moving horizontally from left to right in any given row shows a
consistent and dramatic decrease in both MSE and bias, reinforcing our
finding that performance improves with more proxies. In stark
contrast, Figure~\ref{fig:heatmap_ols} for OLS Screening shows that in
the lower-right quadrant (e.g., $p > 10$, $k > 10$), where regularized
methods thrive, the relative error for OLS remains stubbornly high,
often exceeding 30--40\%. This visually confirms that OLS screening
suffers from severe model selection mistakes in high-dimensional
settings. Furthermore, moving vertically from top to bottom in the
heatmaps reveals the challenge of the `many true surrogates' regime.
For all methods, MSE and bias tend to increase as $p$ grows for a
fixed $k$, confirming the findings from Figure~\ref{fig:te_fixk_varyp}
on a broader scale.

\section{Empirical Application}\label{sec:empirical}

\begin{figure}
    \centering
    \includegraphics[width=\linewidth]{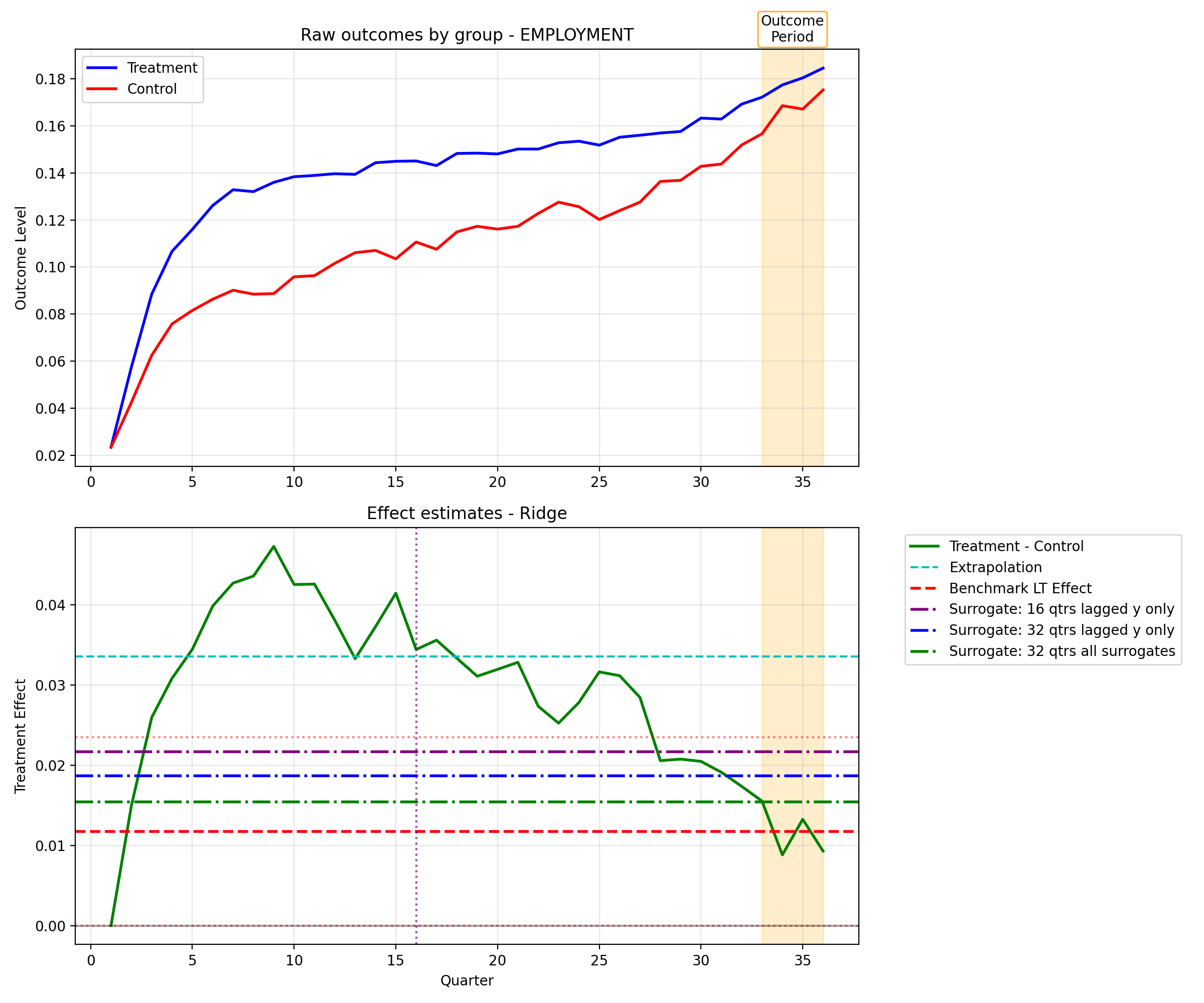}
    \caption{Raw differences (top panel), and naive and surrogate estimates (bottom panel)}
    \label{fig:ridge_gain}
\end{figure}

Next, we apply our estimators to  the classic California GAIN
experiments. The GAIN experiment studied in
\textcite{hotz2006evaluating} was an experiment in a job assistance
program implemented in California in the 1980s to help welfare
recipients find work. We have data on employment, earnings, and
receipt of aid over the first thirty-six quarters (9 years) after
random assignment. We focus on the employment rate (binary indicator
of employment for each individual for each quarter) as the primary
outcome, and focus on the employment rate in the last year as the
primary outcome of interest. We visualize the raw employment rates in
the top panel, the simple difference in means for each quarter and
estimates from multiple estimators in the bottom panel in
Figure~\ref{fig:ridge_gain}. The red dashed line is the true treatment
effect (and the 95\% confidence interval is indicated by the dotted
light red lines). The blue dashed line shows the estimates from
naively extrapolating the average treatment effect in the first 32
periods to the long-run, and is outside the confidence interval for
the long-term effect.

\begin{figure}[t!]
    \centering

    \begin{subfigure}[t]{\textwidth}
    \centering
    \includegraphics[height=4.2in]{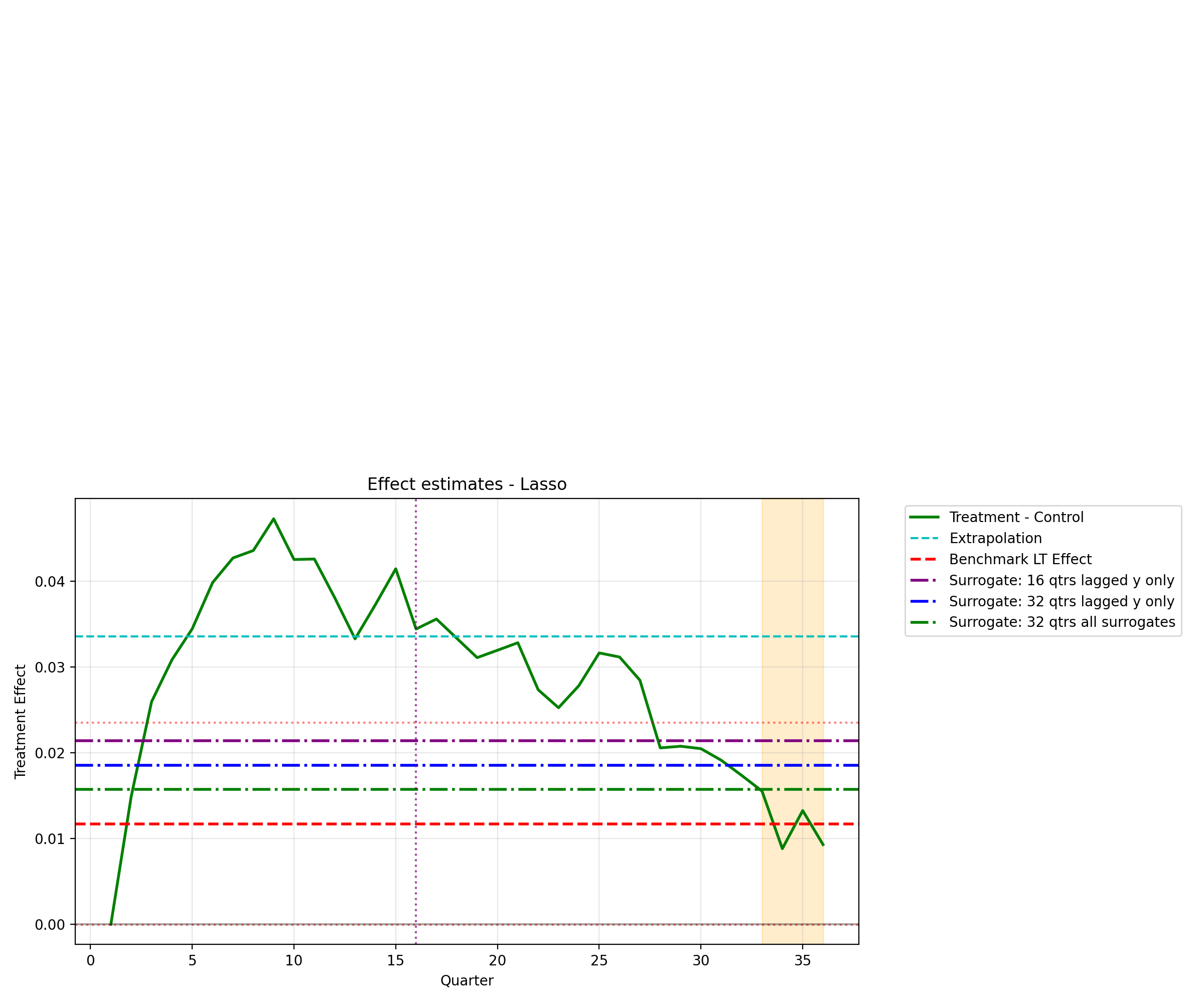}
    \caption{Naive and surrogate estimates from lasso}
    \label{fig:lasso_gain}
    \end{subfigure}
    
    \begin{subfigure}[t]{\textwidth}
     \centering
    \includegraphics[height=4.2in]{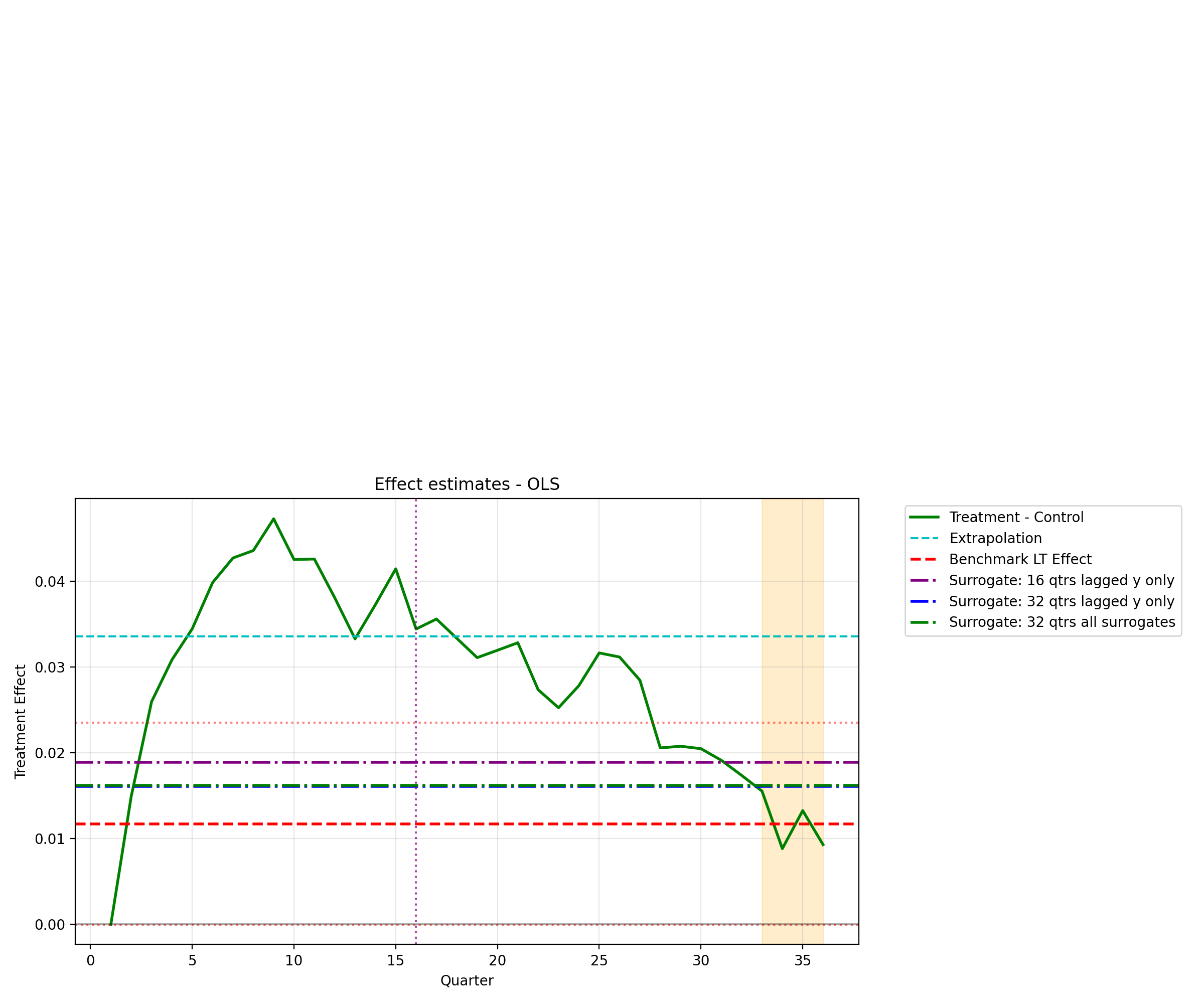}
    \caption{Naive and surrogate estimates from OLS}
    \label{fig:ols_gain}
    \end{subfigure}

 \caption{Naive and Surrogate estimates (Top: Lasso, Bottom: OLS)}
 
\end{figure}

Figure~\ref{fig:ridge_gain} shows the ridge surrogate index gets
closer to the true estimate as we add more periods (16 quarters to 32:
purple vs blue) and more measurements (lagged employment to lagged
employment, earnings, and aid service use: blue vs green). In stark
contrast, Figure~\ref{fig:ols_gain} shows that OLS performance does
not improve with the inclusion of additional surrogates (blue to
green).

In summary, we find that the ridge estimator benefits from the
inclusion of multiple noisy proxies, which bears out the theoretical
and numerical results in previous sections.

\FloatBarrier

\section{Conclusion}

This paper provides both theoretical and empirical evidence in favor
of using regularized regression methods for the linear surrogate index
estimator in a setting with many noisy proxies of low-dimensional
unobserved surrogates. We believe this is an accurate description of
the long-term causal inference problem in many settings in the tech
industry where firms seek to estimate returns on a long-term metric
such as revenue and have access to a large and growing number of
correlated and noisy short-run metrics.

While our numerical experiments evaluate several regularized
estimators, our theoretical analysis focuses on Ridge regression due
to its analytical tractability. It is important to distinguish the
mechanisms by which these methods operate. Both Ridge and Lasso are
defined as solutions to a penalized optimization problem, regularizing
the model by shrinking the norm of the coefficient vector $\alpha$.
Our analysis shows how this L2-shrinkage translates into a simple,
interpretable attenuation bias on the final treatment effect estimate.
In contrast, methods like Partial Least Squares (PLS) regularize the
model via dimensionality reduction. PLS is defined by an iterative
algorithm that projects the high-dimensional proxies P onto a small
number of latent components chosen to maximize covariance with the
outcome Y. A theoretical analysis of PLS would therefore require
analyzing the properties of this algorithmic procedure, including the
approximation error from using a low-rank model and the statistical
uncertainty in the estimated components themselves. Our finding that
Ridge, PLS, and Lasso all outperform naive screening suggests that the
benefit of regularization is a robust phenomenon. We focus on Ridge as
it provides the clearest theoretical insight into the bias-variance
trade-off in this dense surrogate setting.

\printbibliography

\pagebreak

\appendix

\section{Additional Derivations}
\subsection{Deriving the coefficients in the scalar setting}\label{app:alpha_solve}

First, we find the elements of these matrices:
\begin{itemize}
    \item $\Var(P_1) = \Var(S + \varepsilon_1) = \sigma_S^2 + \sigma_1^2$
    \item $\Var(P_2) = \Var(S + \varepsilon_2) = \sigma_S^2 + \sigma_2^2$
    \item $\Cov(P_1, P_2) = \Cov(S + \varepsilon_1, S + \varepsilon_2) = \Cov(S,S) = \sigma_S^2$
    \item $\Cov(P_1, Y) = \Cov(S + \varepsilon_1, \beta S + \eta) = \beta\Cov(S,S) = \beta\sigma_S^2$
    \item $\Cov(P_2, Y) = \Cov(S + \varepsilon_2, \beta S + \eta) = \beta\Cov(S,S) = \beta\sigma_S^2$
\end{itemize}
The system of equations is:
\begin{equation*}
    \begin{pmatrix}
        \sigma_S^2 + \sigma_1^2 & \sigma_S^2 \\
        \sigma_S^2 & \sigma_S^2 + \sigma_2^2
    \end{pmatrix}
    \begin{pmatrix}
        \alpha_1 \\ \alpha_2
    \end{pmatrix}
    =
    \begin{pmatrix}
        \beta\sigma_S^2 \\ \beta\sigma_S^2
    \end{pmatrix}
\end{equation*}
Solving this system yields the expressions for $\alpha_1$ and $\alpha_2$. 

\section{Derivations for Asymptotic Analysis}\label{app:asym_deriv}

\subsection{Derivation of the Asymptotic Bias}
\label{app:bias_derivation}
We start from the general bias expression in Section \ref{eqn:bias_gen} and introduce the simplifying homoskedasticity assumptions: $\Sigma_S = \sigma^2_S I_p$, $\Sigma_E = \sigma^2_E I_k$, and the SVD of the loading matrix $L = U D V^\tra$.
\begin{align}
\text{Bias}_\lambda  
    & = \gamma^\tra  \left[ 
        VDU^\tra  
        (UDV^\tra \Sigma_S VDU^\tra +
            \Sigma_E + \lambda I_k)^{-1}
            UDV^\tra \Sigma_S - I_p
    \right] \beta  \nonumber \\
    & = \gamma^\tra  \left[ 
        VDU^\tra  
        (UDV^\tra (\sigma_S^2 I_p) VDU^\tra +
            (\sigma^2_E + \lambda) I_k)^{-1}
            UDV^\tra \sigma_S^2 - I_p
    \right] \beta & \text{Plug in $\Sigma_E, \Sigma_S$} \\
    & = \gamma^\tra  \left[ 
        VDU^\tra  
        \left(
            \sigma_S^2 U D^2 U^\tra +
            (\sigma^2_E + \lambda) I_k
        \right)^{-1}
        UDV^\tra \sigma_S^2 - I_p
    \right] \beta & \text{Orthogonality of V} \label{eqn:bias_svd_app}
\end{align}
Let $A = \sigma_S^2 U D^2 U^\tra + (\sigma^2_E + \lambda) I_k$. To invert $A$, we use the identity $I_k = UU^\tra + U_{\bot} U_{\bot}^\tra$, where the columns of $U_\bot$ form an orthonormal basis for the orthogonal complement of the column space of $U$ \parencite[Ch 7]{murphy2022probabilistic}.
\begin{align*}
    A & = \sigma_S^2 U D^2 U^\tra + (\sigma^2_E + \lambda) (UU^\tra +  U_{\bot} U_{\bot}^\tra) \\
    & = U(\sigma_S^2 D^2 + (\sigma_E^2 + \lambda) I_p) U^\tra + (\sigma_E^2 + \lambda) U_{\bot} U_{\bot}^\tra
\end{align*}
This matrix is block-diagonal in the basis $[U, U_\bot]$, so its inverse is:
$$
A^{-1} = U(\sigma_S^2 D^2 + (\sigma_E^2 + \lambda) I_p)^{-1} U^\tra + \frac{1}{\sigma_E^2 + \lambda} U_{\bot} U_{\bot}^\tra
$$
Plugging this into equation \eqref{eqn:bias_svd_app} and using $U^\tra U = I_p$ and $U^\tra U_{\bot} = 0$:
\begin{align}
\text{Bias}_\lambda
    & = \gamma^\tra  \left[ 
        VDU^\tra  
        \left(
         U(\sigma_S^2 D^2 + (\sigma_E^2 + \lambda) I_p)^{-1} U^\tra + \frac{1}{\sigma_E^2 + \lambda} U_{\bot} U_{\bot}^\tra       
        \right)
        UDV^\tra \sigma_S^2 - I_p
    \right] \beta \nonumber \\
    & = \gamma^\tra \left[
        VD (\sigma_S^2 D^2 + (\sigma_E^2 + \lambda) I_p)^{-1} D \sigma_S^2 V^\tra - I_p
        \right] \beta 
\end{align}
Using the balanced loadings assumption, $L^\tra L = V D^2 V^\tra = c I_p$, which implies $D^2=cI_p$ and $D=\sqrt{c}I_p$:
\begin{align}
\text{Bias}_\lambda 
    & = \gamma^\tra \left[ V \left( \frac{c \sigma_S^2}{c \sigma_S^2 + \sigma_E^2 + \lambda} I_p \right) V^\tra - I_p \right] \beta \nonumber \\
    & = \gamma^\tra \left[ \left( \frac{c \sigma_S^2}{c \sigma_S^2 + \sigma_E^2 + \lambda} - 1 \right) I_p \right] \beta \tag{Since $VV^\tra = I_p$} \\
    & = - \left( \frac{\sigma_E^2 + \lambda}{c \sigma_S^2 + \sigma_E^2 + \lambda} \right) \gamma^\tra \beta = - \left( \frac{\sigma_E^2 + \lambda}{c \sigma_S^2 + \sigma_E^2 + \lambda} \right) \tau^{*} \nonumber
\end{align}

\subsection{Derivation of the Asymptotic Variance}
\label{app:var_derivation}
The asymptotic variance of $\hat{\tau}_{\lambda}$ is approximately $$\frac{1}{n} \vect{\tau}_p^\tra \left( \mathrm{AVar}(\sqrt{n}(\hat{\vect{\alpha}}_\lambda - \vect{\alpha}_\lambda)) \right) \vect{\tau}_p,$$ where we ignore estimation error in the first-stage effects $\tau_p$ as second-order and can abstract from covariances due to $\hat{\tau}_p$ and $\hat{\alpha}_p$ being estimated on separate datasets. The variance of the Ridge coefficients is given by the sandwich formula $\mathrm{AVar}(\sqrt{n}(\hat{\vect{\alpha}}_\lambda)) = \sigma_\eta^2 A_\lambda^{-1} A A_\lambda^{-1}$, where $A = \E[P P^\tra]$ and $A_\lambda = A + \lambda I_k$. We compute the quadratic form $\vect{\tau}_p^\tra A_\lambda^{-1} A A_\lambda^{-1} \vect{\tau}_p$.
Let $\matr{\Lambda}_p = (\sigma_s^2\matr{D}^2 + (\sigma_e^2+\lambda)\matr{I}_p)$. From the bias derivation, we have the intermediate term $\vect{\tau}_p^\tra A_\lambda^{-1} = \vect{\gamma}^\tra\matr{V}\matr{D}^\tra \matr{\Lambda}_p^{-1} \matr{U}^\tra$.
\begin{align*}
    &\vect{\tau}_p^\tra A_\lambda^{-1} A A_\lambda^{-1} \vect{\tau}_p \\
    &= \left( \vect{\gamma}^\tra\matr{V}\matr{D}^\tra \matr{\Lambda}_p^{-1} \matr{U}^\tra \right) A \left( \matr{U} \matr{\Lambda}_p^{-1} \matr{D}\matr{V}^\tra\vect{\gamma} \right) \\
    &= \vect{\gamma}^\tra\matr{V}\matr{D}^\tra \matr{\Lambda}_p^{-1} ( \matr{U}^\tra A \matr{U} ) \matr{\Lambda}_p^{-1} \matr{D}\matr{V}^\tra\vect{\gamma} \\
    &= \vect{\gamma}^\tra\matr{V}\matr{D}^\tra \matr{\Lambda}_p^{-1} ( \matr{U}^\tra(\sigma_s^2 \matr{U}\matr{D}^2\matr{U}^\tra + \sigma_e^2\matr{I}_k)\matr{U} ) \matr{\Lambda}_p^{-1} \matr{D}\matr{V}^\tra\vect{\gamma} \\
    &= \vect{\gamma}^\tra\matr{V}\matr{D}^\tra \matr{\Lambda}_p^{-1} (\sigma_s^2\matr{D}^2 + \sigma_e^2\matr{I}_p) \matr{\Lambda}_p^{-1} \matr{D}\matr{V}^\tra\vect{\gamma}
\end{align*}
Using the balanced loading assumption ($D^2 = cI_p$), $\matr{\Lambda}_p$ becomes the scalar matrix $(c\sigma_s^2 + \sigma_e^2 + \lambda)\matr{I}_p$.
\begin{align*}
    ... &= \vect{\gamma}^\tra\matr{V} (\sqrt{c}I_p) \frac{c\sigma_s^2 + \sigma_e^2}{(c\sigma_s^2 + \sigma_e^2 + \lambda)^2}I_p (\sqrt{c}I_p) \matr{V}^\tra\vect{\gamma} \\
    &= c \frac{c\sigma_s^2 + \sigma_e^2}{(c\sigma_s^2 + \sigma_e^2 + \lambda)^2} \vect{\gamma}^\tra (\matr{V}\matr{V}^\tra) \vect{\gamma} = c \frac{c\sigma_s^2 + \sigma_e^2}{(c\sigma_s^2 + \sigma_e^2 + \lambda)^2} ||\vect{\gamma}||^2
\end{align*}
Therefore, the asymptotic variance of $\hat{\tau}_{\lambda}$ is $\Var_{asy}(\hat{\tau}_{\lambda}) = \frac{\sigma_\eta^2}{n} \left( c \cdot ||\vect{\gamma}||^2 \cdot \frac{c\sigma_s^2 + \sigma_e^2}{(c\sigma_s^2 + \sigma_e^2 + \lambda)^2} \right)$.

\subsection{Derivation of MSE and the Optimal Regularization Parameter $\lambda^*$}
\label{app:lambda_derivation}

Combining the squared bias and the asymptotic variance, we arrive at the full AMSE:
\begin{equation}
\boxed{
    \AMSE(\hat{\tau}_{\lambda}) = \underbrace{ \left( \frac{\sigma_e^2 + \lambda}{c\sigma_s^2 + \sigma_e^2 + \lambda} \right)^2 (\tau^*)^2 }_{\text{Squared Bias}} + \underbrace{ \frac{\sigma_\eta^2 c ||\vect{\gamma}||^2}{n} \frac{c\sigma_s^2 + \sigma_e^2}{(c\sigma_s^2 + \sigma_e^2 + \lambda)^2} }_{\text{Variance}}
}
\end{equation}
This expression formalizes the bias-variance trade-off. The bias term is monotonically increasing in $\lambda$, while the variance term is monotonically decreasing in $\lambda$. Minimizing this AMSE with respect to $\lambda$ yields the optimal regularization parameter, $\lambda^*$, providing the theoretical justification for using Ridge regression and cross-validation in this dense surrogate setting.

\paragraph{Optimal Regularization Strength $\lambda^*$}

To find the value of $\lambda$ that minimizes this expression, we differentiate with respect to $\lambda$ and set the result to zero. To simplify the notation during the derivation, let's define the following constants:

\begin{itemize}
    \item $A = c\sigma_s^2 + \sigma_e^2$
    \item $S = \sigma_e^2$
    \item $C_1 = (\tau^*)^2$
    \item $C_2 = \frac{\sigma_\eta^2 c ||\vect{\gamma}||^2 (c\sigma_s^2 + \sigma_e^2)}{n}$
\end{itemize}

The AMSE can now be written as a function of $\lambda$:
\begin{equation}
    \AMSE(\lambda) = \frac{C_1 (S + \lambda)^2 + C_2}{(A + \lambda)^2}
\end{equation}

Using the quotient rule, we compute the derivative $\frac{d}{d\lambda} \AMSE(\lambda)$ and set its numerator to zero:
\begin{align*}
    \frac{d}{d\lambda} \AMSE(\lambda) &= \frac{[2 C_1 (S+\lambda)](A+\lambda)^2 - [C_1(S+\lambda)^2 + C_2][2(A+\lambda)]}{(A+\lambda)^4} = 0 \\
    \implies 2 C_1 (S+\lambda)(A+\lambda)^2 &= 2 [C_1(S+\lambda)^2 + C_2](A+\lambda)
\end{align*}

We can divide by $2(A+\lambda)$, which is non-zero for $\lambda \ge 0$:
\begin{align*}
    C_1 (S+\lambda)(A+\lambda) &= C_1(S+\lambda)^2 + C_2 \\
    C_1(S+\lambda)[(A+\lambda) - (S+\lambda)] &= C_2 \\
    C_1(S+\lambda)(A - S) &= C_2
\end{align*}

This final line represents the optimal balance where the marginal change in squared bias equals the marginal change in variance. We now substitute the original terms back into this equation:
\begin{itemize}
    \item $A - S = (c\sigma_s^2 + \sigma_e^2) - \sigma_e^2 = c\sigma_s^2$
    \item $S = \sigma_e^2$
    \item $C_1 = (\tau^*)^2$
    \item $C_2 = \frac{\sigma_\eta^2 c ||\vect{\gamma}||^2 (c\sigma_s^2 + \sigma_e^2)}{n}$
\end{itemize}

Plugging these in yields:
\begin{equation*}
    (\tau^*)^2 (\sigma_e^2 + \lambda) (c\sigma_s^2) = \frac{\sigma_\eta^2 c ||\vect{\gamma}||^2 (c\sigma_s^2 + \sigma_e^2)}{n}
\end{equation*}

We can now solve for $\lambda$. The term $c$ cancels from both sides:
\begin{align*}
    (\tau^*)^2 (\sigma_e^2 + \lambda) \sigma_s^2 &= \frac{\sigma_\eta^2 ||\vect{\gamma}||^2 (c\sigma_s^2 + \sigma_e^2)}{n} \\
    \sigma_e^2 + \lambda &= \frac{\sigma_\eta^2 ||\vect{\gamma}||^2 (c\sigma_s^2 + \sigma_e^2)}{n (\tau^*)^2 \sigma_s^2}
\end{align*}

This gives the final analytic expression for the optimal regularization parameter, $\lambda^*$:
\begin{equation}
\boxed{
    \lambda^* = \frac{\sigma_\eta^2 ||\vect{\gamma}||^2 (c\sigma_s^2 + \sigma_e^2)}{n (\tau^*)^2 \sigma_s^2} - \sigma_e^2
}
\end{equation}

Since the regularization parameter must be non-negative, the optimal choice is $\lambda_{\text{opt}} = \max(0, \lambda^*)$. This result confirms that the optimal level of regularization depends on underlying properties of the data generating process, such as the sample size, outcome noise, and the alignment of treatment effects, thus justifying the use of data-driven methods like cross-validation in practice.

\section{Additional Simulation Results}

\subsection{Varying both the number of proxies and number of true surrogates}\label{appdx:gridsim}

\newgeometry{left=1cm, right=1cm, top=1cm, bottom=0cm}

\begin{figure}
    \centering
    \includegraphics[width=1\linewidth]{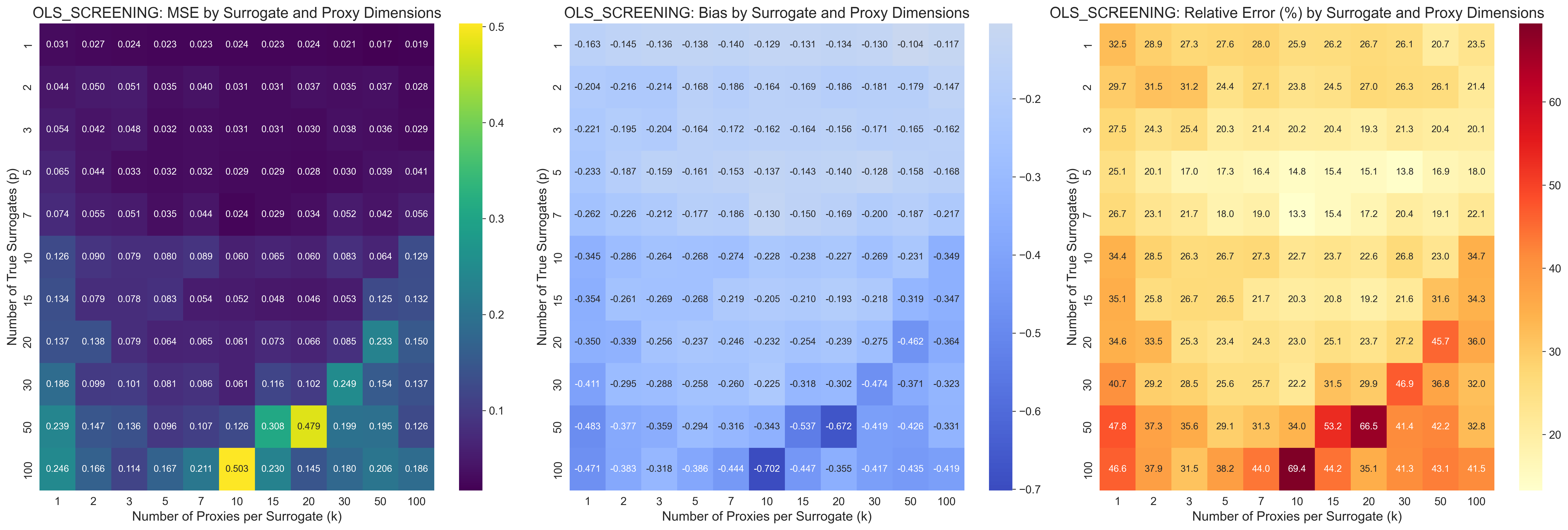}
    \caption{RMSE, Bias, and Bias relative to Oracle for OLS Screening with varying number of true surrogates and proxies}
    \label{fig:heatmap_ols}
\end{figure}

\begin{figure}
    \centering
    \includegraphics[width=1\linewidth]{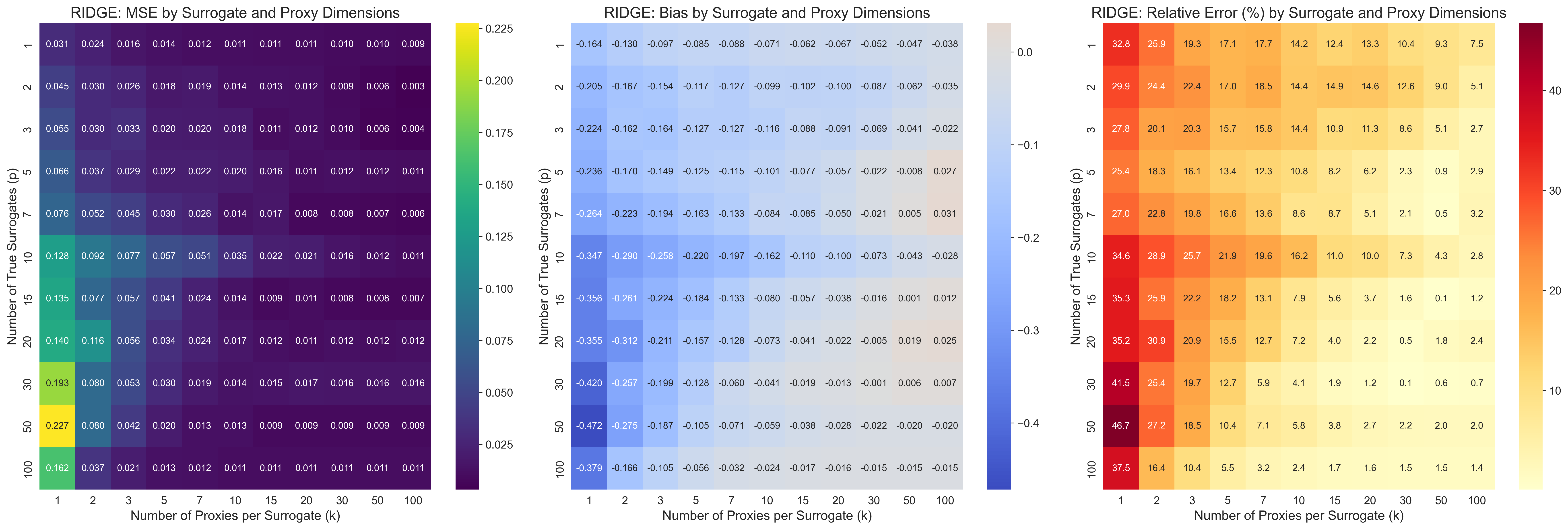}
    \caption{RMSE, Bias, and Bias relative to Oracle for Ridge with varying number of true surrogates and proxies}
    \label{fig:heatmap_ridge}
\end{figure}

\begin{figure}
    \centering
    \includegraphics[width=1\linewidth]{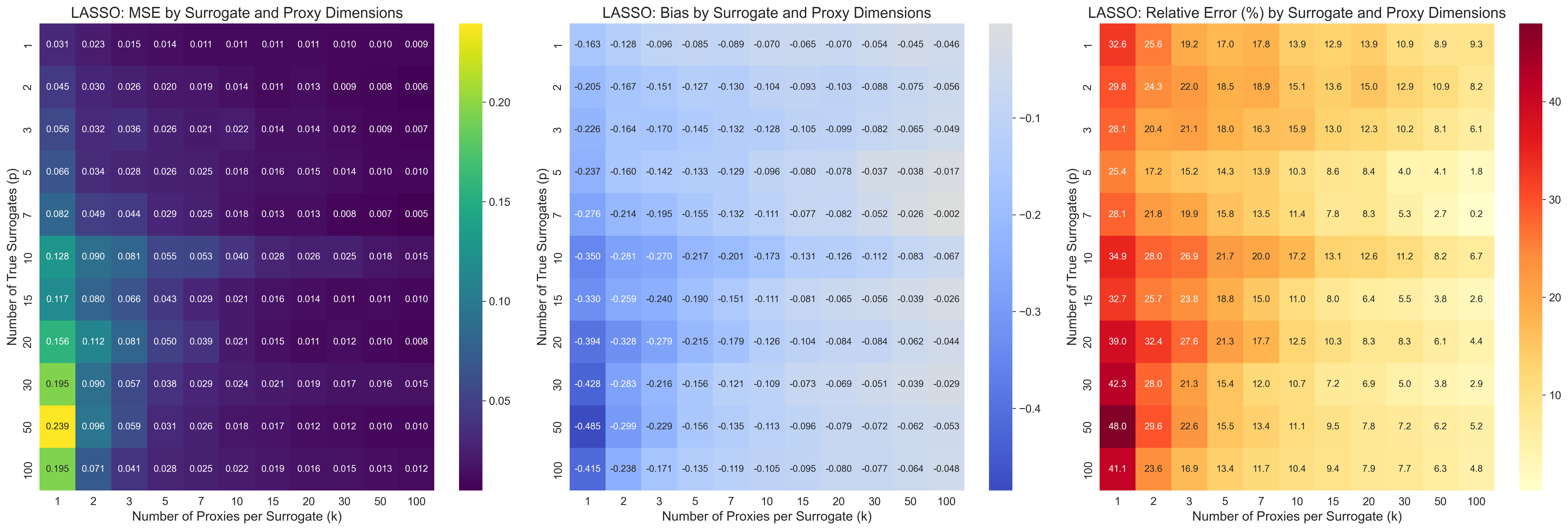}
    \caption{RMSE, Bias, and Bias relative to Oracle for LASSO with varying number of true surrogates and proxies}
    \label{fig:heatmap_lasso}
\end{figure}

\begin{figure}
    \centering
    \includegraphics[width=1\linewidth]{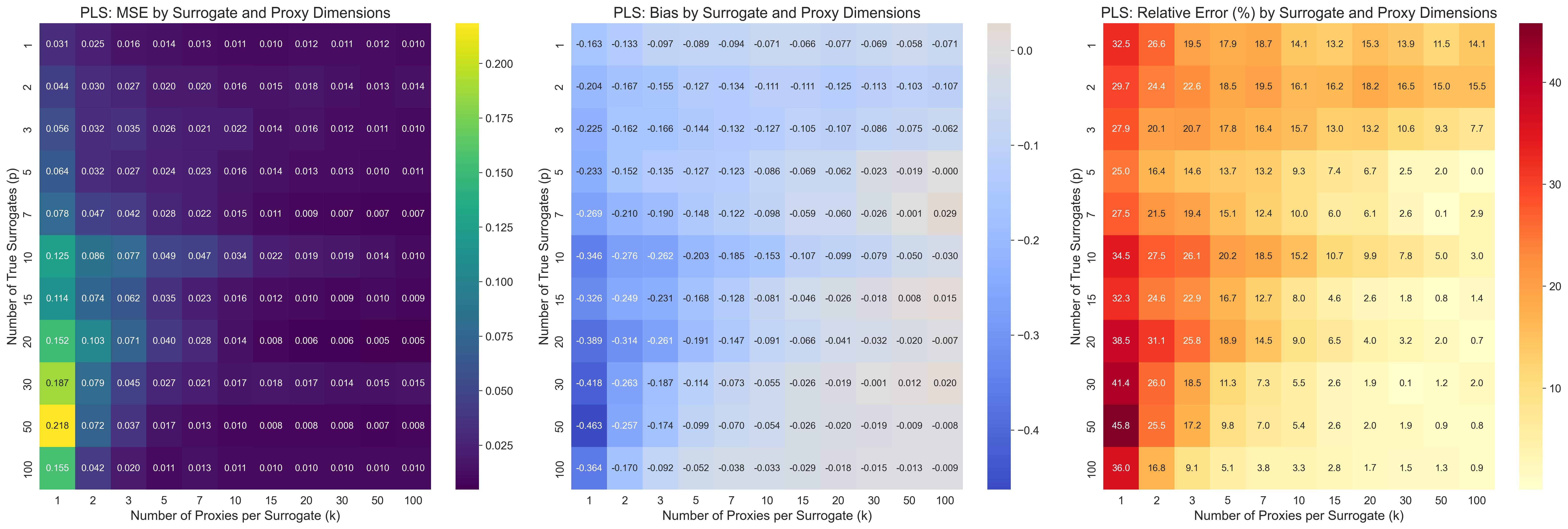}
    \caption{RMSE, Bias, and Bias relative to Oracle for PLS with varying number of true surrogates and proxies}
    \label{fig:heatmap_pls}
\end{figure}

\restoregeometry

\FloatBarrier

\end{document}